\documentclass[12pt,a4paper]{article}
\usepackage{amsmath,amssymb,bm,ascmac,bbm,mathtools}
\usepackage[dvipdfmx, usenames]{color}
\usepackage[dvipdfmx]{graphicx}
\usepackage{xcolor}
\usepackage{here}
\usepackage{authblk}
\usepackage[hang,small,bf]{caption}
\usepackage[subrefformat=parens]{subcaption}
\usepackage{dcolumn}

\newcommand{\mbb}{\mathbb}
\newcommand{\mfrak}{\mathfrak}

\newcommand{\mc}{\mathcal}
\newcommand{\uv}{\text{UV}}
\newcommand{\ir}{\text{IR}}

\begin{document}
\title{Monotonicities of Tanaka-Nakayama flows}
\author{Ken KIKUCHI} 
\affil{Department of Physics, National Taiwan University, Taipei 10617, Taiwan}
\date{}
\maketitle

\begin{abstract}
We prove conformal and global dimensions monotonically decrease under the infinitely many Tanaka-Nakayama renormalization group flows between Virasoro minimal models. The flows also satisfy the half-integer condition.
\end{abstract}

\makeatletter
\renewcommand{\theequation}
{\arabic{section}.\arabic{equation}}
\@addtoreset{equation}{section}
\makeatother

\section{Introduction}
Why we do not observe Higgs bosons in our daily lives? The reason is because they are too heavy for our energy scales. In other words, their lifetime is too short to observe. They affect our daily lives through effective couplings. The idea of effective theory goes back to Kadanoff \cite{K66}. He invented the block-spin transformations. The transformation is defined as follows. Consider a lattice system with spin degrees of freedom (up or down) at each lattice site. One combines some (say nine) lattice sites into a block and define a block-spin variable. The block-spin variable of a block is defined to be up if there are more up spins in the block, and defined to be down if there are more down spins. In this way, one reduces the number of degrees of freedom. (In case of a block with nine spins, the number of degrees of freedom reduces to $1/9$ per block-spin transformation.) The resulting theory is an effective theory. In the effective theory, we no longer see the original spin variables but only the block-spin variables are available. In this way, we gradually drop unnecessary details while keep important physics. By repeating the block-spin transformations, one arrives a convenient effective theory with much fewer degrees of freedom.

The original idea of block-spin transformation was imported to high-energy physics by Wilson \cite{WK73,W75}. In the Wilsonian renormalization group (RG), one divides integration variables in path integral into ``heavy'' modes and ''light'' modes. By integrating out ``heavy'' modes, one arrives at an effective field theory. This picture teaches us two physical intuitions. First, from the analogy with block-spin transformations, we expect effective field theories have less and less degrees of freedom. In fact, the intuition has been proven in various spacetime dimensions; $c$-theorem \cite{cthm} (or effective $c$-theorem \cite{CDR17} for non-unitary flows with $PT$-symmetry) in two dimensions, $F$-theorem \cite{JKPS11,CDFKS12} in three dimensions, $a$-theorem \cite{C88,KS11,K11} in four dimensions, and $a$-theorem \cite{CDY15,CDI15} in six dimensions. The second physical intuition the picture tells us is that we expect effective field theories have lighter and lighter degrees of freedom (or more precisely operators). The intuition as well has been proven in special cases \cite{KKRCFT2,KKfree,KKWZW}.

Recently, it was also observed that RG flows can be understood thermodynamically \cite{KKfree}. More concretely, just like other physical phenomena, RG flows are also realized to minimize free energy. A fusion category symmetry $\mc C$ of a theory contributes free energy by $F\ni T\ln D(\mc C)$ at temperature $T$ (defined as length $1/T$ of the Euclidean time compactified to a circle). (The entropy is called the topological entanglement entropy \cite{KP05,LW05}.) Here, $D^2(\mc C)$ is the global dimension of $\mc C$.\footnote{The topological entanglement entropy was proposed for unitary theories. In these theories, a square root $D(\mc C)$ of the global dimension given by the modular $S$-matrix as $D(\mc C)=1/S_{11}$ is positive. Thus, the contribution $T\ln D(\mc C)$ is guaranteed to be real. On the other hand, in non-unitary theories, the square root $D(\mc C)=1/S_{11}$ can be negative. For example, in the Lee-Yang model, we have $D(\mc C)=-\sqrt{\frac{5-\sqrt5}2}$. While we do not know how to define topological entanglement entropy for negative $D(\mc C)$, a natural candidate is to take its real part, $T\mfrak R\ln D(\mc C)=T\ln|D(\mc C)|$. Note its similarity with the definition of the supersymmetric R\'enyi entropy \cite{NY13}. With this in mind, for non-unitary theories, we expect the global dimension $D^2(\mc C)=|D(\mc C)|^2$ to decrease monotonically under RG flows.} From the principle of free energy minimization, we also expect the contribution to free energy gets smaller and smaller as we lower energies. The monotonicity has also been proven in special cases \cite{KKfree,KKWZW}.

Last year, it was observed \cite{TN24} based on \cite{Z87,A92,L92,DDT00,KNST22,LMMT22,LMMT23} that the RG flows between Virasoro minimal models could be special classes of more general RG flows $M(kq+I,q)+\phi_{1,2k+1}\to M(kq-I,q)$ for an integer $I=1,2,\dots$ and a fixed natural number $q$ so that $gcd(kq\pm I,q)=1$.\footnote{We follow the notation of \cite{TN24}.} Another parameter $k$ could be integer or half-integer. This raises an immediate question: Are the Tanaka-Nakayama flows also monotonic? We answer this question positively by proving the\\

\textbf{Theorem.} \textit{The conformal dimensions monotonically decrease under the Tanaka-Nakayama renormalization group flows $M(kq+I,q)+\phi_{1,2k+1}\to M(kq-I,q)$:}
\begin{equation}
    h^{M(kq+I,q)}_j\ge h^{M(kq-I,q)}_j\label{monotonich}
\end{equation}
\textit{for preserved symmetry objects $j$ with Kac index $(r,1)$. Furthermore, the conformal dimensions also satisfy the half-integer condition:}
\begin{equation}
    h^{M(kq+I,q)}_j+h^{M(kq-I,q)}_j\in\frac12\mbb Z.\label{halfinteger}
\end{equation}
\textit{For $0<\frac q{kq+I}<\frac q{kq-I}\le\frac12$, the global dimensions also decrease monotonically:}
\begin{equation}
    D_{M(kq+I,q)}^2>D_{M(kq-I,q)}^2.\label{monotonicD2}
\end{equation}\\

The second statement may need an explanation. It was conjectured in \cite{KKWZW} and shown in \cite{KK24Witt} that when an RG flow to a rational conformal field theory (RCFT), called rational RG flow, admits an RG domain wall/interface \cite{G12}, the conformal dimensions of surviving symmetry objects should add up to a half-integer. While Gaiotto constructed RG domain wall/interface explicitly only for rational RG flows between unitary minimal models, he claimed the construction should work more generally. At least, the physical picture of relevant deformation only in half of the spacetime seems to work as well in the Tanaka-Nakayama flows. If they indeed admit RG domain wall/interface, they should satisfy the half-integer condition by the theorem in \cite{KK24Witt}. The theorem above proves the flows in fact satisfy the condition.\footnote{The fact can be interpreted in two ways of the same coin. If a reader believes the Tanaka-Nakayama flows are correct, then the fact supports the flows admit RG domain wall/interface. If one believes the flows admit RG domain wall/interface, the fact gives further support of the Tanaka-Nakayama flows.} We prove the theorem in the next section.

Before presenting a proof, let us summarize all known monotonicities of (rational) RG flows:
\begin{itemize}
    \item $c$-theorem $c^\uv>c^\ir$ and its generalizations.
    \item Spin constraint $S^\uv_j\supset S^\ir_j$ \cite{KKSUSY} for scalar relevant operators.
    \item Monotonicity of conformal dimensions $h^\uv_j\ge h^\ir_j$. This is still conjectural beyond the special cases.
    \item Monotonicity of global dimensions $D^2(\mc C^\uv)>D^2(\mc C^\ir)$. This is still conjectural beyond the special cases. There are also some `counterexamples' for flows from non-unitary to unitary RCFTs.
\end{itemize}
The spin constraint (and the matching of quantum dimensions provided by monoidal functor \cite{KKARG}) was the main support of the Tanaka-Nakayama flow. The spin constraint claims the following. Given a symmetry object $j\in\mc C^\uv$, one can insert the object parallel to the time-direction. The object intersect a time slice, and hence modifies the Hilbert space. The modified Hilbert space $\mc H_j$ is called the twisted Hilbert space. Its operators have specific spins $S^\uv_j$ called spin content. For an RG flow triggered by a scalar operator, rotation symmetry is preserved. Hence, its quantum number, spins are conserved. Thus, one expects $S^\uv_j,S^\ir_j$ for a preserved symmetry object $j$ to be the same. More precisely, however, some heavy operators in the twisted UV Hilbert space $\mc H_j$ may be ``integrated out'' during RG flows, and its spin would be absent in the twisted IR Hilbert space $\mc H_j$. Therefore, IR spin contents $S^\ir_j$ should be a subset of UV spin contents $S^\uv_j$, $S^\uv_j\supset S^\ir_j$ for a preserved symmetry object $j$. Since the first two monotonicities have been checked in \cite{TN24}, we only prove the last two monotonicities.

\section{Proof}
In the Tanaka-Nakayama flows $M(kq+I,q)+\phi_{1,2k+1}\to M(kq-I,q)$, Verlinde lines \cite{V88} $L_{r,1}$ with $r=1,\dots,q-1$ are preserved for integer $k$, and $r=2l+1\le q-1$ for $l\in\mbb N$ are preserved for half-integer $k$. The preserved symmetry objects flow as follows:
\begin{equation}
\begin{array}{cccccc}
M(kq+I,q):&L_{1,1}&\cdots&L_{r,1}&\cdots&L_{q-1,1}\\
&\downarrow&\cdots&\downarrow&\cdots&\downarrow\\
M(kq-I,q):&L_{1,1}&\cdots&L_{r,1}&\cdots&L_{q-1,1}
\end{array},\label{preservedsym}
\end{equation}
or an appropriate modification for the last object for half-integer $k$.

In order to prove the two monotonicities, all one has to know are the two formulae. For a Virasoro minimal model $M(p,p')$ with coprime natural numbers $p,p'$, conformal dimensions and modular $S$-matrices are given by (for example, see the yellow book \cite{FMS})
\begin{equation}
\begin{split}
    h_{r,s}&=\frac{(pr-p's)^2-(p-p')^2}{4pp'},\\
    S_{(r,s),(r',s')}&=(-1)^{1+rs'+sr'}\sqrt{\frac8{pp'}}\sin\left(\pi\frac p{p'}rr'\right)\sin\left(\pi\frac{p'}pss'\right)
\end{split}\label{hS}
\end{equation}
for Kac indices in the Kac table
\[ E_{p,p'}:=\{(r,s)|1\le r\le p'\&1\le s\le p\}/\sim. \]
The equivalence of two indices are defined by
\[ (r,s)\sim(p'-r,p-s). \]
The modular $S$-matrix gives (a square root of) the global dimension
\begin{equation}
    D_{M(p,p')}=\frac1{S_{(1,1),(1,1)}}=-\frac1{\sqrt{\frac8{pp'}}\sin\pi\frac p{p'}\sin\pi\frac{p'}p}.\label{D}
\end{equation}
We will prove the monotonicities and the half-integer condition separately for integer and half-integer $k$.

\subsection{Integer $k$}
From the formula (\ref{hS}), the difference of conformal dimensions are given by
\begin{equation}
    h^{M(kq+I,q)}_{r,1}-h^{M(kq-I,q)}_{r,1}=\frac{I(r^2-1)}{2q}\quad(r=1,2,\dots,q-1),\label{hdifference}
\end{equation}
which is nonnegative. (The computation shows the equality in (\ref{monotonich}) is satisfied only for the unit object $L_{1,1}$.) Furthermore, we see they satisfy the half-integer condition
\begin{equation}
    h^{M(kq+I,q)}_{r,1}+h^{M(kq-I,q)}_{r,1}=\frac12(r-1)(kr+k-2)\in\frac12\mbb Z.\label{halfintegercond}
\end{equation}
The ratio of global dimensions is given by
\begin{equation}
    \frac{D_{M(kq+I,q)}^2}{D_{M(kq-I,q)}^2}=\frac{kq+I}{kq-I}\frac{\sin^2\frac{\pi q}{kq-I}}{\sin^2\frac{\pi q}{kq+I}}.\label{Dratio}
\end{equation}
Since $kq-I>0$, the first factor is larger than one. The second factor needs a care. Since $\sin^2x$ monotonically increases for $0\le x\le\frac\pi2$, the second factor is also larger than one if $0\le\frac q{kq-I},\frac q{kq+I}\le\frac12$. This proves the monotonicity $D_{M(kq+I,q)}^2>D_{M(kq-I,q)}^2$ for those $(k,q,I)$.

For the other $(k,q,I)$, case-by-case analysis is needed. This is so for a reason; for some parameters, we find an \textit{increase} of global dimensions. For example, $(k,q,I)=(2,4,3)$, or $M(11,4)\to M(5,4)$ has
\[ 13.3\approx D_{M(11,4)}^2<D_{M(5,4)}^2\approx14.5. \]
Another `counterexample' is $(k,q,I)=(2,5,4)$, or $M(14,5)\to M(6,5)$. This flow has
\[ 31.2\approx D_{M(14,5)}^2<D_{M(6,5)}^2\approx43.4. \]
From these examples, it seems that global dimensions could increase under rational RG flows from non-unitary UV theory to unitary IR theory.

\subsection{Half-integer $k$}
When $k$ is a half-integer, $r$ does not run all $1,2,\dots,q-1$, but only odd integers $r=2l+1$ with $l=0,1,\dots$ so that $r\le q-1$. However, the equation (\ref{hdifference}) holds for half-integer $k$ as well for $r=2l+1$. This proves the monotonicity of conformal dimensions. Similarly, their sum (\ref{halfintegercond}) also holds with $r=2l+1$. In this case, the RHS becomes
\[ (\text{RHS})=l\Big(2k(l+1)-2\Big)\in\mbb Z. \]
Hence, the half-integer condition is satisfied. Finally, the ratio of global dimensions is still given by (\ref{Dratio}). Therefore, for $0<\frac q{kq+I}<\frac q{kq-I}\le\frac12$, the global dimensions decrease monotonically.


\appendix
\setcounter{section}{0}
\renewcommand{\thesection}{\Alph{section}}
\setcounter{equation}{0}
\renewcommand{\theequation}{\Alph{section}.\arabic{equation}}


\end{document}